**Vapor-liquid-solid growth of highly-mismatched semiconductor nanowires with high-fidelity van der Waals layer stacking**

Edy Cardona[1,2]*, Matthew K. Horton[1,2], Daniel Paulo-Wach[1,2], Anthony C. Salazar[1,2], Andre Palacios Duran[2], James Chavez[1,2], Shaul Aloni[3], Junqiao Wu[1,2], Oscar D. Dubon[1,2]**

[1]Department of Materials Science and Engineering, University of California, Berkeley, California 94720, USA

[2]Materials Sciences Division, Lawrence Berkeley National Laboratory, Berkeley, California 94720, USA

[3]Molecular Foundry, Lawrence Berkeley National Laboratory, Berkeley, California 94720, USA

Nanobelts, nanoribbons, and other quasi-one-dimensional nanostructures formed from layered, so-called, van der Waals semiconductors have garnered much attention due to their high-performance, tunable optoelectronic properties. For layered alloys made from the gallium monochalcogenides GaS, GaSe, and GaTe, near-continuous tuning of the energy bandgap across the full composition range has been achieved in $GaSe_{1-x}S_x$ and $GaSe_{1-x}Te_x$ alloys. Gold-catalyzed vapor-liquid-solid (VLS) growth of these alloys yields predominantly nanobelts, nanoribbons and other nanostructures for which the fast crystal growth front consists of layer edges in contact with the catalyst. We demonstrate that in the S-rich, $GaS_{1-x}Te_x$ system – unlike $GaSe_{1-x}S_x$ and $GaSe_{1-x}Te_x$ – the Au-catalyzed VLS process yields van der Waals nanowires for which the fast growth direction is normal to the layers. The high mismatch between S and Te leads to extraordinary bowing of the $GaS_{1-x}Te_x$ alloy's energy bandgap, decreasing by at least 0.6 eV for x as small as 0.03. Calculations using density functional theory confirm the significant decrease in bandgap in S-rich $GaS_{1-x}Te_x$. The nanowires can exceed fifty micrometers in length, consisting of tens of thousands of van der Waals-bonded layers with triangular or hexagonal cross-sections of uniform dimensions along the length of the nanowire. We propose that the low solubility of Te in GaS results in an enhancement in Te coverage around the Au catalyst-nanowire interface, confining the catalyst to the chalcogen-terminated basal plane (rather than the edges) and thereby enabling layer-by-layer, c-axis growth.

*Email: ecardona@berkeley.edu
**Email: oddubon@berkeley.edu

Layered, two-dimensional (2D) semiconductors possess desirable properties for applications in next-generation optoelectronic devices among other applications[1,2]. Alloying is a common approach used to precisely engineer the properties of materials, particularly their energy bandgap, for specific applications. Several alloying studies have reported full bandgap tunability of layered materials, such as the transition metal dichalcogenide (TMD) alloys $MoS_{2(1-x)}Se_{2x}$[3–7], $MoSe_{2(1-x)}Te_{2x}$[8], $WS_{2(1-x)}Se_{2x}$[9,10], $WSe_{2(1-x)}Te_{2x}$[11], $Mo_{1-x}W_xS_2$[12], and $Mo_{1-x}W_xSe_2$[13,14]. In these materials, the bandgap follows an approximately linear dependence with composition because electronegativity and atomic radius are well-matched properties between the constituent binary compounds.

Highly-mismatched semiconductor alloys (HMAs) are more difficult to synthesize due to the larger lattice strains and solubility limits imposed by their highly mismatched constituents. Yet, HMAs have important technological applications because their electronic structures near the band edges can be tuned significantly with a few percent of alloying and because the bandgap range of the alloy system can be much larger than the bandgap range bounded by the constituents' endpoint compounds – that is, they display giant bandgap bowing. Traditional highly-mismatched systems, such as $GaAs_{1-x}N_x$[15] and $ZnO_{1-x}S_x$[16], are typically thin-film materials and have been studied primarily in covalently or mixed covalently/ionically bonded semiconductors. The electronic and morphological properties of HMAs that possess van der Waals, inter-layer bonding have received limited attention thus far.

The vapor-liquid-solid (VLS) growth mechanism offers a unique path to study the morphological features of van der Waals, highly-mismatched systems for two reasons. First, VLS can yield structures with unique growth morphologies. Examples of these novel structures include Ag-catalyzed GaS nanowires[17]; screw-dislocation-mediated, helical GeS nanostructures[18–20]; and a variety of other complex, quasi-one-dimensional morphologies[21]. Moreover, nanoscale structures are known to accommodate a greater amount of lattice mismatch compared to bulk crystals[22]. Fonseca *et al.*[23] and Cai *et al.*[24] synthesized small $GaSe_{1-x}Te_x$ crystals across the entire composition range, whereas Cammassel *et al.*[25] had previously observed a miscibility gap in bulk crystals of the same alloy system.

The gallium monochalcogenide family – composed of the group III-VI binary compounds GaS, GaSe, and GaTe – is ideal to study the band engineering of semiconductor alloys through crystal phase selection in addition to compositional control. For example, $GaSe_{1-x}S_x$[26–33] crystals exist as both β and ε hexagonal polytypes while $GaSe_{1-x}Te_x$[23,24] crystals exist in both hexagonal and monoclinic phases, each crystal phase possessing a different range of bandgap energies. In this family of materials, GaS and GaTe possess the greatest relative property mismatch; GaS is typically an *n*-type semiconductor with an indirect fundamental bandgap of 2.59 eV and a direct bandgap of 3.05 eV at 77K[34] while GaTe is a *p*-type semiconductor with a direct bandgap of 1.78 eV at 4.3 K[5,36]. GaS has a thermodynamically stable hexagonal crystal structure in the β polytype (space group *P63/mmc*, lattice parameters a = 0.36 nm and c = 1.55 nm)[34] with each unit cell containing two layers, each of which is four atoms, S-Ga-Ga-S, stacked along the c-axis[37,38]. In contrast, GaTe has a thermodynamically stable monoclinic crystal structure (space group *C2/m*, lattice parameters a = 1.74 nm, b = 0.41 nm c = 1.05, β = 104.44°)[39–41] with two different Ga-Ga configurations: one third parallel to the layer and two thirds perpendicular to the layer[42]. Sulfur has an electronegativity of 2.58, which is 0.48 larger than the electronegativity of tellurium. Finally, tellurium's atomic radius is 140 pm, which is 40 pm larger than the atomic radius of sulfur. Due to the high degree of mismatch between the properties of GaS and GaTe binary compounds, the $GaS_{1-x}Te_x$ alloy system is expected to be an alloy that offers the advantages of both a layered material and a highly mismatched alloy. However, $GaS_{1-x}Te_x$ synthesis has not been reported thus far.

Here, we present the synthesis of van der Waals, highly-mismatched $GaS_{1-x}Te_x$ nanowires by gold-catalyzed VLS growth. The incorporation of Te into the growth process changes the fast-growth direction from being in the plane of the layers for pure GaS (nanobelts) to perpendicular to the layers (c-axis direction) in the case of $GaS_{1-x}Te_x$ alloys (nanowires). Room-temperature micro-photoluminescence spectra are characterized by prominent peaks centered at ~1.42 eV and ~1.56 eV, indicating a significant decrease of the fundamental (indirect) bandgap of GaS via dilute alloying with GaTe (i.e., x ≤ 0.13 in $GaS_{1-x}Te_x$), typical of a highly-mismatched alloy. Calculations using density functional theory (DFT) further support a picture in which the $GaS_{1-x}Te_x$ alloy system displays strong bandgap bowing in the sulfur-rich regime of composition.

**Results and Discussion**

**Nanowire morphology.** $GaS_{1-x}Te_x$ nanowires were synthesized using a gold-catalyzed vapor transport method, as shown in Figure 1a. A typical alloy growth yields nanostructures of various morphologies as shown by scanning electron microscopy (SEM) in Figures 1b. Notably, nanowires have grown, and these can extend many tens of micrometers in length. The VLS mechanism is the dominant growth process for these nanowires[43] as evidenced by the solidified gold catalyst droplets present at the tip of the nanowires in Figures 1f, 2a, and 2d, and by the high aspect ratio of their morphologies.

$GaS_{1-x}Te_x$ nanostructures with nominal tellurium compositions of 0 < x ≤ 0.13 possess different morphologies, such as nanoribbons (seen for x<0.5) and nanowires. However, this work focuses on nanowires that meet the following criteria: near 1:1 stoichiometric composition between gallium and chalcogen species, continuously straight edges parallel to the fast-growth direction of the nanowire (i.e., the c-axis of the nanowire), smooth surfaces perpendicular to the c-axis of the nanowire, and a uniform cross-section, as shown in Figure 1c-e.

The layered nature and high-fidelity stacking of these van der Waals nanowires are evident in Figure 1g, in which a segment of the nanowire was cleaved by a mechanical shear force when landing on top of another, narrower nanowire during the transfer process. Nanowire cross-sections vary from triangular to regular hexagonal, as shown in Figures 1c-e. The edge length of the triangular/hexagonal cross-section (normal to the growth direction) ranges from approximately 100 nm to up to 2 µm.

Selected area diffraction (SAED) analysis indicates that GaS$_{1-x}$Te$_x$ nanowires are monocrystalline. The diffraction pattern in Figure 2b was indexed in accordance with a [210] zone axis of the hexagonal crystal system. The fast-growth direction of the nanowires is thus confirmed to be along [001]. The lattice parameters of an alloy nanowire with nominal composition x = 0.11 were calculated to be a = 0.365 nm and c = 1.580 nm. These values reflect a 1.60% and 1.96% expansion of a and c, respectively, compared to GaS. Such expansion is consistent with the incorporation of Te, which has a larger atomic radius (140 pm) than S (100 pm). Finally, Ga, S, and Te appear to be randomly and homogeneously distributed along the length the nanowire as shown by the energy-dispersive X-ray spectroscopy (EDS) maps in Figure 2d. However, these measurements cannot eliminate the possibility of local inhomogeneity on the atomic and few-unit-cells scale within individual layers.

Further evidence that Te is incorporated into the chalcogen sublattice of GaS$_{1-x}$Te$_x$ nanowires is found in the Raman spectra measured as a function of composition, presented in Figure 2f. The positions of the three GaS Raman-active modes are indicated by dashed, vertical lines. With the addition of Te, these Raman modes red shift, consistent with the substitution of S by Te. Mode $A_{1g}^1$ splits into two peaks for x > 0.04. Such splitting has been observed for the GaS$_{1-x}$Se$_x$ alloy system[44,45] for which the splitting of the $A_{1g}$ modes is attributed to the co-existence of multiple (stacking) polytypes within the crystal. The co-existence of two stacking types in GaS$_{1-x}$Se$_x$ is consistent with optical measurements that indicate the presence of ε and β polytypes[31]. We therefore interpret the splitting of the $A_{1g}$ modes in S-rich GaS$_{1-x}$Te$_x$ to the coexistence of the β and another polytype, possibly ε.

We attribute the unique growth behavior of GaS$_{1-x}$Te$_x$ nanowires compared to all other Au-catalyzed growth of Ga-monochalcogenides to solubility kinetics. The Au catalyst is a liquid at the growth temperature, and Te is more soluble in the liquid than in the growing GaS$_{1-x}$Te$_x$ crystal[46]. Therefore, as Te is expelled from the catalyst, some of it incorporates into the GaS$_{1-x}$Te$_x$ bulk – determined by the solubility of Te in GaS at the growth temperature – while the excess Te remains on the basal and/or diffuses to the its edges, stabilizing the Au-catalyst droplet on the basal plane during growth. The stabilization of gold on the basal plane (depicted schematically in Figure 1c and observed by SEM in Figure 1f) results in the alloy nanowire having a fast-growth direction parallel to the c-axis of the basal plane. The GaS$_{1-x}$Te$_x$ alloy nanowire growth kinetics differ from those previously reported for Ga-monochalcogenide nanobelts[47]. For such nanobelts the gold catalyst droplet is not located exclusively on the basal plane but on the unterminated edges of the basal plane, leading to a nanobelt fast-growth direction perpendicular to the c-axis, or fast edge growth. We note that the Au-catalyzed growth of van der Waals nanowires has not been observed in the GaS$_{1-x}$Se$_x$ and GaSe$_{1-}$ $_x$Te$_x$ systems both of which display a high miscibility between the corresponding endpoint compounds.

A perhaps related growth mechanism has been proposed for the case of the Au-catalyzed, vapor-solid-solid (VSS) growth of Bi$_2$Se$_3$ nanowires[48]. Specifically, reaction at the Au-Bi$_2$Se$_3$ interface of selenium species from the vapor with bismuth expelled from the solid catalyst leads to nanowire growth if the catalyst is located on the basal plane of a growing Bi$_2$Se$_3$ nanostructure, such as a nanoribbon.

**Luminescence properties.** The luminescence of GaS$_{1-x}$Te$_x$ alloy nanowires was studied by micro-photoluminescence (micro-PL) spectroscopy using 488 nm (2.54 eV) and 660 nm (1.88 eV) laser lines and by cathodoluminescence (CL) spectroscopy. Room-temperature PL spectra measured with the 488-nm laser line are presented in Figure 3a and summarized in Figure 3c. The spectra from nanowires in the compositional range of 0.3 ≤ x ≤ 0.13 display two broad, major peaks centered at approximately ~1.42 eV and ~1.56 eV as well as one low-intensity peak at 2.24 eV. Similarly, the PL spectra that were measured at room temperature with a 660-nm excitation laser line display two, broad main peaks centered at ~1.37 eV and ~1.46 eV; an example is presented Figure 3d. The difference in the PL peak energies between spectra measured with 488-nm and 660-nm laser lines may be due to local laser heating[49,50]. This is consistent with the higher laser power required when using the 660-nm laser line to acquire PL spectra, which nonetheless have a lower signal-to-noise ratio compared to the spectra acquired with the 488-nm line. The positions of measured peaks are relatively insensitive to the narrow composition range (x) over which nanowires were synthesized successfully.

A sample CL spectrum measured at 85 K is presented in Figure 3b and compared to a PL spectrum of the same nanowire, which has a composition of x=0.11. The major CL peak is centered at 2.27.eV while a broad signal of lower peak intensity is centered at 1.86 eV. These peaks do not match the major PL peaks observed, indicating that the dominant recombination mechanisms in the PL and CL may differ due to the significantly higher excitation rates present in a typical CL measurement. As previously noted, the PL spectrum does display a weak peak that is centered near 2.24 eV. This higher-energy PL peak is located at an energy slightly below that of the major CL peak[34,50]. We note that the CL intensity at room temperature was significantly lower, precluding direct comparison of the peak positions between PL and CL.

**DFT calculations and band structure model for GaS$_{1-x}$Te$_x$ alloy nanowires.** Calculations using density functional theory (DFT) of the GaS$_{1-x}$Te$_x$ alloy system were performed with the PBEsol and HSE06 exchange-correlation functionals[51], with necessary structural relaxations and density of states obtained using PBEsol, and band gap corrections applied with HSE06. The density-of-states calculations for different alloy compositions are displayed in Figure 4a-c. As more Te replaces S, the valence band restructures by hybridization with Te ($p$) states. The stronger impact of Te states on the valence band compared to the conduction band in the S-rich limit is expected in a highly mismatch semiconductor alloy due to the higher electronegativity and smaller size of S compared to Te. The plot of energy versus composition in Figure 4d reveals that the bandgap energy of pure GaS decreases with the incorporation of Te by nearly 0.50 eV at x=0.06 for both HSE06 and PBEsol calculations. The band bowing observed in the DFT calculations is consistent with the behavior of a highly mismatched alloy whereby significant band bowing as a result of an anticrossing interaction between Te states and the GaS host is expected.

The maximum solubility limit at different compositions was calculated as shown in Figure 4F. From these calculations, the maximum Te content in a nanowire grown at 750 °C is x=0.16, which is much lower than the solubility of Te in liquid Au at this temperature[46]. Consistent with these calculations, the measured nanowire compositions across multiple growths at this temperature do not exceed x=0.13.

We propose a band structure model that explains the intriguing luminescence behavior of $GaSe_{1-x}Te_x$. At dilute concentrations of Te in GaS, a valence-band anticrossing interaction between the Te states and the valence band states of GaS causes significant restructuring of the valence band[52,53]. Such band restructuring results in significant bowing of the energy bandgap with even low levels of incorporation of Te. Similar behavior has been reported in conventional semiconductors such as ZnO containing small amounts of S – larger, less electronegative S replacing O[16]. In this case as well as that of $GaS_{1-x}Te_x$, the energy bandgap decreases abruptly upon initial alloying, resulting in the reduction of the fundamental bandgap. In this framework, we assign the CL peak centered at 2.27 eV and the small PL peak at 2.24 eV to an interband transition across the direct bandgap. (The slight difference in the PL and CL peak positions may be caused by the difference in measurement temperature.) Therefore, the substitution of small quantities of S with Te reduces the direct bandgap by approximately 0.8 eV (from 3.05 eV to 2.27 eV). We attribute the broad CL signal centered at 1.85 eV to defect-mediated interband transitions.

Because the incorporation of small amounts of Te into GaS represents primarily a valence-band anticrossing interaction, one expects the direct and indirect bandgaps in $GaS_{1-x}Te_x$ to decrease by approximately similar amounts. In the example of $GaS_{0.89}Te_{0.11}$, a 0.8 eV reduction of the direct bandgap would suggest an indirect bandgap for $GaS_{0.89}Te_{0.11}$ of around 1.8 eV. We therefore attribute the major PL peaks below 1.6 eV to interband transitions across the indirect gap. In GaS, luminescence from recombination by indirect excitons has been observed at room-temperature[54]. The co-existence of luminescence due to direct and indirect interband transitions has been observed in $MoS_2$[55].

The presence of two PL peaks below 1.6 eV can be attributed to two stacking polytypes, β and ε, within the $GaS_{1-x}Te_x$ nanowires. This possibility is based on the co-existence of polytypes reported in the $GaS_{1-x}Se_x$ alloy system[30]. In the case of $GaSe_{1-x}S_x$, a room-temperature indirect bandgap of 2.53 eV was inferred for β-GaS by photomodulated reflectance spectroscopy[30]. In the same study, a room-temperature PL peak centered at 2.43 eV was assigned to the ε polytype of GaS contained within a mainly β-polytype crystal, suggesting a difference of ~0.1 eV between indirect bandgaps of the two polytypes; a small or negligible difference in bandgap is expected from DFT calculations. A schematic depiction of the proposed model is presented in Figure 4f.

At this time, we cannot discount the possibility that the PL peaks below 1.6 eV are due to defect-mediated recombination. If this is the case, then the likely process for defect PL involves a conduction-band-to-defect transition. This follows from the band anticrossing framework whereby conduction-band restructuring due to alloying is not significant in the S-rich end of the compositional range (i.e., small x). Therefore, a transition from the conduction band to a defect level in the bandgap should be relatively insensitive to alloy composition. Indeed, the experimentally observed PL peak positions display a weak compositional dependence (Figure 3 a and c). While the observed PL peaks may be defect-related, our measurements still provide insight about the magnitude of the indirect band gap. Given that a conduction-band-to-defect transition requires photoexcitation across the bandgap and that PL peaks persist even when spectra are acquired using a 660-nm laser line (Figure 3d), the magnitude of the indirect bandgap is very likely less than 1.88 eV (660 nm) – that is, at least 0.6 eV below the bandgap of GaS.

$GaS_{1-x}Te_x$ nanowires are intriguing materials both fundamentally and technologically. These van der Waals nanowires offer a unique opportunity to study through-layer electronic and thermal transport as well as mechanical behavior. Further, the high accessibility of the interlayer region over the entire length of the nanowire facilitates the investigation of intercalants and their influence on a range of properties while catalysis and chemical sensing may be facilitated by the nanowire's surface, which is composed of layer edges. The significant bandgap reduction and environmental stability of $GaS_{1-x}Te_x$ merit further investigation for applications in photodetector technology[56–62].

In summary, $GaS_{1-x}Te_x$ alloy nanowires were synthesized for the first time. Unlike all other Au-catalyzed VLS-grown nanostructures in the Ga-monochalcogenide family, the fast-growth direction of $GaS_{1-x}Te_x$ nanowires is parallel to the c-axis. The difference in the fast-growth direction between layer edges and c-axis originates from the stabilization of the Au catalyst on the basal plane by Te atoms due to the limited solubility of Te in GaS at the growth temperatures. The luminescence spectra support a picture in which a valence-band anticrossing interaction induces an extraordinary bowing of the energy bandgap in S-rich $GaS_{1-x}Te_x$. The application of the VLS process to other highly mismatched alloys made from van der Waals semiconductors opens unique opportunities for both bandgap and shape engineering of quasi-one-dimensional nanostructures.

**Acknowledgments**

All work was supported by the Office of Science, Office of Basic Energy Sciences, of the US Department of Energy under Contract DE-AC02-05CH11231. Materials growth and optical characterization were performed in the Electronic Materials Program, and correlative microscopy including CL, TEM and EDS were performed at the Molecular Foundry. The authors recognize that this work was performed at the Berkeley Lab and UC Berkeley, which sit on the territory of xučyun (Huichin), the ancestral and unceded land of the Chochenyo speaking Ohlone people, and that every member of the Berkeley community continues to benefit from the use and occupation of this land. The authors gratefully acknowledge Dr. Jerry Tersoff for insightful communications about nanowire growth mechanisms and Dr. Zakaria Al-Balushi for valuable discussions.

**Methods**

**Nanostructure Synthesis.** Nanostructures were synthesized using a gold-catalyzed, vapor-liquid-solid (VLS) growth mechanism. The substrate was a silicon chip (*p*-type, <100> orientation) with dimensions 7 mm by 15 mm. The substrate was first sonicated in isopropyl alcohol for 15 minutes to remove debris from the cleavage process. After drying the substrate with nitrogen gas, the substrate was cleaned in oxygen plasma for five minutes to remove hydrocarbons from the surface. Finally, 5 nm of gold were deposited on the substrate using e-beam evaporation at a rate of 0.05 nm/s.

The source materials were 50 mg of GaS and 50 mg of GaTe powders (American Elements, 99.999% purity). For synthesis of alloy nanowires, each powder was placed in separate cylindrical alumina crucibles (height = 12 mm, outer diameter = 8 mm). For the synthesis of GaS nanobelts, only GaS powder was used. The substrate and crucibles were placed inside a cylindrical quartz furnace tube (length = 600 mm, inner diameter = 20.5 mm) and the tube was loaded into a single-zone furnace (Lindberg/Blue M furnace).

The growth recipe consisted of three steps. In the first step, the furnace was heated to 70°C and the furnace tube was purged with Argon gas (Praxair, 99.999% purity) at 200 sccm for 30 minutes to remove water vapor. In the second step, the furnace was heated to 1000°C and the gas flow was reduced to 100 sccm for 30 minutes. In this step, the temperatures of the GaS powder was 1000°C. The temperature of GaS was chosen based on the thermal decomposition temperatures reported for gallium monochalcogenides[63]. The temperature of the GaTe powder was optimized to obtain x up to 0.13 when combined with an appropriate substrate temperature. For example, for a growth with average composition x = 0.10, the GaTe powder was placed at a location where the furnace temperature was 643 °C while the region on the substrate showing the highest nanowire density after growth was positioned at a furnace temperature of 750 °C (Figure 1a). These temperatures were determined from detailed thermal profiling along the entire length of the furnace. Importantly, the temperature of the substrate was very sensitive to positioning in the furnace; consequently, the temperature gradient across the growth region – as high as 120 °C/cm – enabled the harvesting of nanowires of different compositions from a single growth run. In the final step, the gas flow was lowered to 10 sccm to minimize nanowire growth as the furnace gradually cooled down to room temperature. The minimal gas flow was necessary to prevent back flow from the potassium hydroxide neutralizing solution at the outflow of the furnace. The furnace tube was kept at atmospheric pressure during all steps.

**Characterization.** Grown nanostructures were mechanically transferred to a handle substrate for correlated microscopy, either a 100 nm $SiO_2$ on Si chip patterned with enumerated gold markers or a SiN TEM grid. The mechanical transfer process consisted of manually aligning the surfaces of the as-grown substrate and handle substrate and applying pressure to transfer wires. This facile method was efficient in transferring a large number of wires to the handle substrate. The morphology,size and composition of individual nanostructures were measured by SEM (FEI Quanta 3D FEG) equipped with an energy dispersive spectrometer (Oxford) prior to optical characterization. Individual nanowires of known compositions were characterized by micro Raman and PL to probe their crystal quality and optical transitions. . The experimental set up of PL and Raman consisted of a 488 nm laser with 100 mW maximum power output as the excitation source, a silicon-based CCD as the detector, a 1200 gr/mm grating for PL, and a 2400 gr/mm grating for Raman. The laser power used for Raman and PL measurements were 5% and 0.05%, respectively, to minimize damage to the nanowire and signal from the handle substrate. Structure and composition of individual nanowires were studied via Transmission electron microscopy was performed on a JEOL 2100F microscope operated at 200kV equipped with energy dispersive x-ray spectrometer (Oxford instruments Ultim Max). Cathodoluminescence spectroscopy was performed on home built instrument based on an Zeiss Gemini Supra 55 VP-SEM , equipped with light collecting parabolic mirror (1.3 sr solid angle), Acton SP2300i spectrometer and Andor Newton EMCCD (970-UVB).

**Density Functional Theory (DFT) Calculations.** DFT calculations were performed using VASP[64,65] following recipes provided by the atomate code[66]. Standard atomate settings were used, except that geometry optimizations were performed with a stricter force tolerance and using the PBEsol[51] density functional approximation, since PBEsol improves on the description of bulk solids and has satisfactory performance[67] for geometry of layered materials without the need for additional empirical corrections. Initial structures were taken from the Materials Project as templates with beta-GaS (mp-2507) for the beta-GaS calculation and epsilon-GaSe (mp-1572) for the epsilon-GaS calculation with S substituted. For intermediate compositions it was assumed that Te would randomly incorporate on the S sub-lattice, and so special quasi-random structures were generated to represent each intermediate alloy composition[68]. Densities of states were generated for all systems but full band structures were only calculated for the end-points due to band folding for the larger supercells required for the intermediate alloy compositions. Since local functionals are known to systematically under-estimate the band gap, additional calculations were performed using the hybrid HSE06 functional which is known to provide more accurate band gaps[69] and a scissor correction then applied to the PBEsol densities of states and band structures as appropriate, and visualized using sumo[70]. Estimates of solubility were obtained by the use of the GQCA method[71-73] and a complete enumeration of all possible supercells containing 32 atoms.

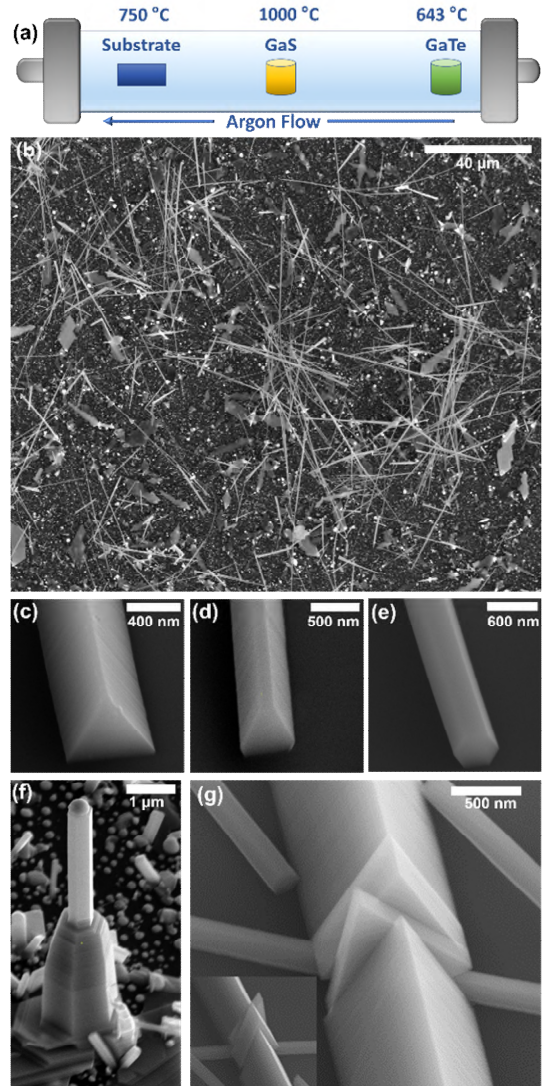

Figure 1. (a) Single-zone furnace set-up used for growth. The temperatures indicated were used to achieve the growth shown in (b). (b) Scanning electron micrograph of a grown sample. The image was captured after the transfer process with nanowires lying flat, revealing their lengths. (c-e) Scanning electron micrographs of transferred nanowires with different cross-sectional morphologies: triangular, irregular hexagonal, and regular hexagonal. (f) Scanning electron micrograph of short hexagonal nanowire that is growing from the facet of a larger crystal. The nanowire is capped by the catalyst. (g) Scanning electron micrograph of the section of a large transferred nanowire with triangular cross section that has been partitioned, or sliced, by a smaller nanowire. The inset in the lower left shows the same nanowire from the opposite perspective. Note that the partitions consist of flat, cleaved surfaces arising from the stacked, layered nature of nanowires. Also, the smaller nanowire near the top left section of the micrograph has an hexagonal cross sections and, like other nanowires in (c-e), is terminated by a flat (cleaved) surface as well.

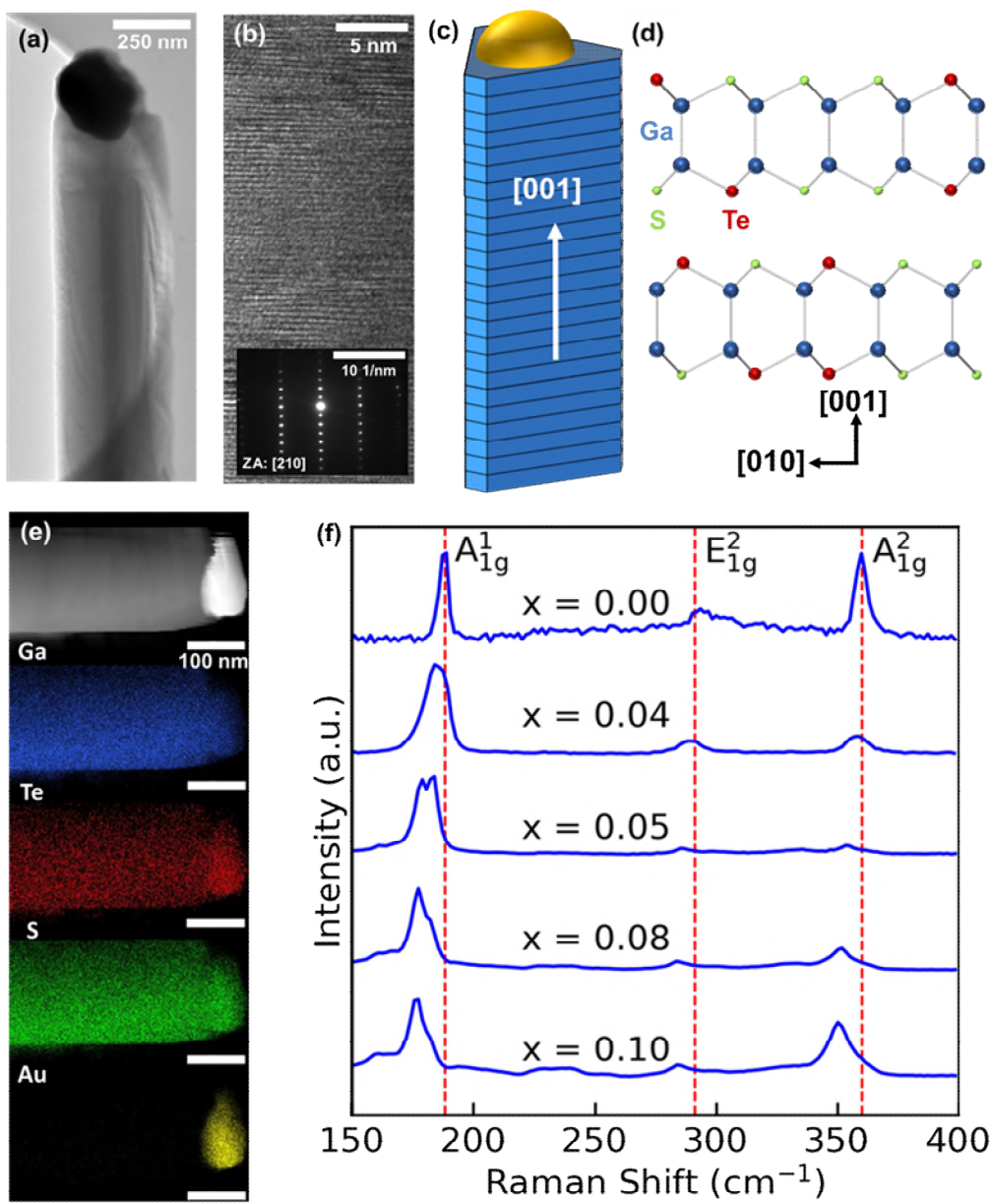

Figure 2. (a) Transmission electron micrograph of a $GaS_{1-x}Te_x$ nanowire showing the catalyst. (b) High-resolution transmission electron micrograph and electron diffraction of the nanowire in (a). (c) Schematic representation of the nanowire along with (d) atomic model of bilayers. (e) Scanning transmission electron microscopy image and energy dispersive spectra showing the distributions of Ga, Te, S, and Au for a $GaS_{0.89}Te_{0.11}$ nanowire. Ga, Te, and S are uniformly distributed throughout while Au is located only at the catalyst end. (f) Micro-Raman spectra of $GaS_{1-x}Te_x$ nanowires with different x (0.00 to 0.10). The redshift of the major peaks is consistent with the incorporation of Te in the GaS matrix.

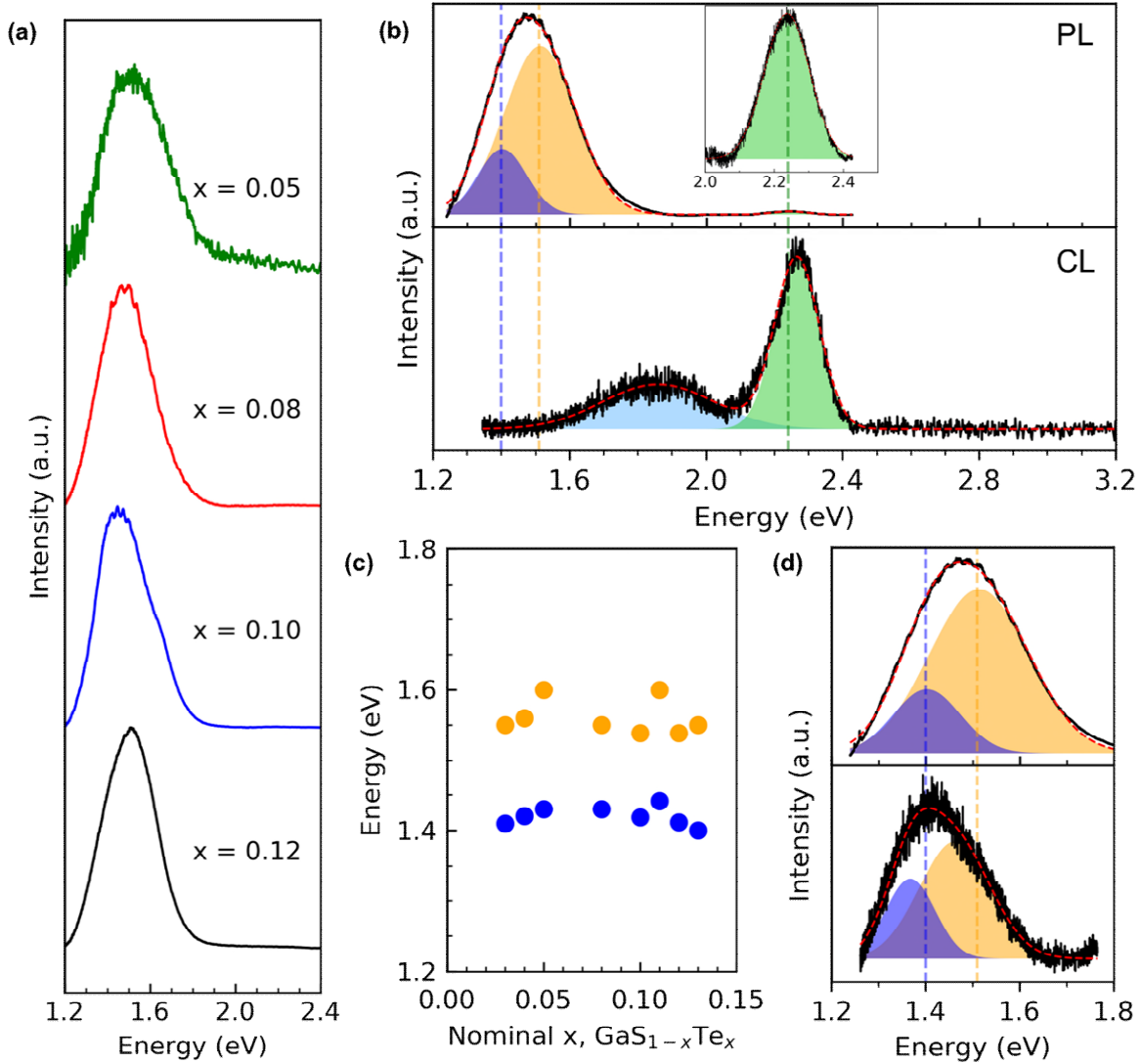

Figure 3. (a) Room-temperature photoluminescence spectra of GaS$_{1-x}$Te$_x$ nanowires with different x up to 0.12 measured with a 488-nm laser line. The strongest luminescence intensity between 1.2 eV and 1.8 eV is characterized by two peaks. (b) Lower panel. Cathodoluminescence spectrum of GaS$_{1-x}$Te$_x$ nanowire with x=0.11. The spectrum is characterized by a peak near 2.27 eV and a broad signal centered around 1.86 eV. Upper panel. Photoluminescence spectrum of same nanowire. Inset shows a magnified region of the spectrum between 2.0 and 2.5 eV and the presence of a peak centered near 2.24 eV (green). Note that the stronger PL signal at the lower energy is not visible in the CL spectrum. (c) PL peak energies versus x obtained from (a). The two peaks are relatively insensitive to composition over the narrow, nominal composition range. (d) Upper panel. PL spectrum of GaS$_{1-x}$Te$_x$ (x=0.11) collected with a 488 nm excitation-laser line. The spectrum is characterized by two peaks centered at 1.40 eV and 1.51 eV. Lower panel. PL spectrum form the same wire measured with a 660 nm excitation-laser line. Similarly, two peaks are identified and centered at 1.37 eV and 1.46 eV.

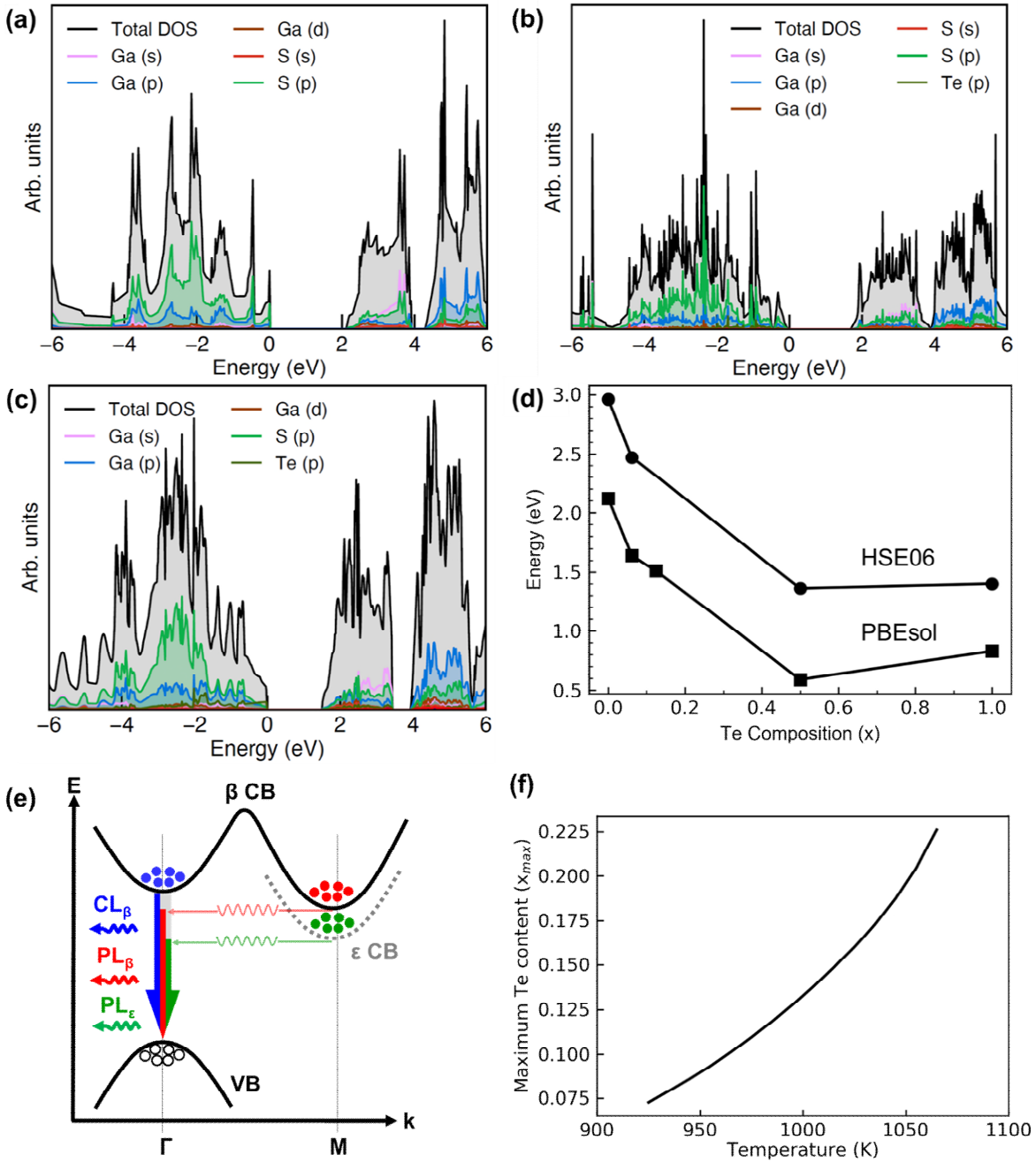

Figure 4. (a) DFT calculation of density of states (DOS) for pure GaS. (b) A single representative DFT calculation of DOS for GaTe:(GaS+GaTe) with 1:16 (x=0.625) (c) A single representative DFT calculation of DOS for GaTe:(GaS+GaTe) with 1:8 (x=0.125) (d) Calculated bandgap using two different DFT approximations. Note the large bandgap bowing on the S-rich side, which is consistent with experiments. (e) Schematic of possible transitions that explain luminescence peaks. The large CL peak at 2.23 eV results from direct bandgap recombination. The PL peaks between ~1.4 eV and ~1.6 eV correspond to recombination across the indirect bandgap. (f) Calculated maximum solubility ($x_{max}$) of Te in GaS in the temperature range of the substrate during growth.